\batchmode
\makeatletter
\def\input@path{{/Users/axelaraneda/Desktop/Research/fBM/}}
\makeatother
\documentclass[english]{article}
\usepackage{lmodern}
\usepackage[T1]{fontenc}
\usepackage[latin9]{inputenc}
\usepackage[a4paper]{geometry}
\usepackage{amsmath}
\usepackage{amsthm}
\usepackage{amssymb}
\usepackage{graphicx}
\usepackage[numbers,numbers,sort&compress]{natbib}

\makeatletter
\newcommand{\lyxaddress}[1]{
	\par {\raggedright #1
	\vspace{1.4em}
	\noindent\par}
}

\usepackage{hyperref}
\date{}

\usepackage{todonotes}
\usepackage{bbm}
\usepackage{fontawesome5}
\usepackage{orcidlink}

\@ifundefined{showcaptionsetup}{}{%
 \PassOptionsToPackage{caption=false}{subfig}}
\usepackage{subfig}
\makeatother

\usepackage{babel}
\begin{document}
\title{\textbf{Price modelling under}\linebreak{}
\textbf{ generalized fractional Brownian motion}}
\author{Axel A.~Araneda\orcidlink{0000-0003-4436-7974}\thanks{Email: \protect\href{mailto:axelaraneda@mail.muni.cz}{\texttt{axelaraneda@mail.muni.cz}}}}
\maketitle

\lyxaddress{\begin{center}
\vspace{-2em} Institute of Financial Complex Systems \\ Department
of Finance\\ Masaryk University\\ 602 00 Brno, Czech Republic.
\par\end{center}}

\begin{center}
\vspace{-1em} This version: \today \vspace{2em}
\par\end{center}
\begin{abstract}
The Generalized fractional Brownian motion (gfBm) is a stochastic
process that acts as a generalization for both fractional, sub-fractional,
and standard Brownian motion. Here we study its use as the main driver
for price fluctuations, replacing the standard Brownian Brownian motion
in the well-known Black-Scholes model. By the derivation of the generalized
fractional Ito\textquoteright s lemma and the related effective Fokker-Planck
equation, we discuss its application to both the option pricing problem
valuing European options, and the computation of Value-at-Risk and
Expected Shortfall. Moreover, the option prices are computed for a
CEV-type model driven by gfBm. 


\vspace{1em}
\end{abstract}

\section{Introduction}

The Black-Scholes model \citep{black1973pricing}, in shortly BS,
is generally categorized as the cornerstone of financial engineering.
By means of Geometric Brownian motion and Delta-hedging arguments,
the model provides a valuation formula solving the partial differential
equation which rules the vanilla option pricing. 

However, some ``stylized facts'' query the BS assumptions. One of
them is the lack of memory (or Markovian property) due to the use
of a standard Brownian motion (Bm) in the price fluctuation modeling.
One attempt to correct it, is the replacement of the Bm by a fractional
Brownian motion (fBm) \citep{mandelbrot1968fractional}, a Gaussian
process characterized by self-similarity, Holder paths, stationary
increments, and long (short) range dependence for $H>1/2\,(H<1/2)$.
The parameter $H\in(0,1)$ is called the Hurst exponent and for $H=1/2$
reduces the fBm to a Bm. An explicit close-form option price solution
for the fractional BS model is given in refs. \citep{necula2002option,hu2003fractional}
followed by a lot of research outputs that consider the analytical
valuation of financial derivatives with underlying assets driven by
fBm. Recent developments include, e.g., the pricing of Asian options
\citep{ahmadian2020pricing}, the European option pricing under the
constant elasticity of variance (CEV) model \citep{araneda2020fractional},
default risk derivatives as Credit Default Swaps \citep{araneda2022credit}
or vulnerable options \citep{Cheng2023}, and the consideration of
jumps or Fuzzy theory \citep{Zhang2021}.

Bojdecki et al. \citep{bojdecki2004sub} introduce a new stochastic
process called sub-fractional Brownian motion (sfBm), which retains
the most properties of fBm but differs in some key issues related
to its non-overlapping increments: non-sta\-tio\-na\-ri\-ty and weakly
covariance (with higher decay rate). More details about this process
are described in \citep{tudor2007some}. Some applications of sfBm
in finance have been published recently. For example, Xu and Zhou
\citep{xu2019pricing} addressed the pricing of perpetual American
options, Wang and its coauthors \citep{wang2021pricing} studied the
pricing of geometric Asian power options, and Ji et al. \citep{Ji2022}
the barrier option valuation in a jump environment. Outside of the
BS environment, we could list, among others, the sub-fractional Poisson
volatility model \citep{wang2021closed}, the sub-fractional Vasicek
model \citep{tao2023asian}, the sub-fractional CEV \citep{araneda2021sub}
which was derived and empirically tested through real option data
finding both an improvement with respect to classical CEV and the
BS types (under classical, fractional and sub-fractional diffusions),
and the capability to capture the option prices temporal structure
under different maturities.

Alternatively, Zili \citep{zili2017generalized} proposed a new Gaussian
process called generalized fractional Brownian motion (gfBm in shorthand
notation), defined as a linear combination of a two-sided fBm, which
serves as a generalization for both Bm, fBm, and sfBm. This process
is featured with the ability to control the level of correlation between
the increments and the ductility to arise stationary and non-stationary
increments. A review of its main properties can be found in refs.
\citep{zili2017generalized,zili2018generalized}.

This manuscript aims to use the gfBm for financial modelling and its
application in the computation of risk-measures and the pricing of
European options. The rest of the paper is organized in the following
way. First, we describe some helpful properties and auxiliary about
gfBm, including the generalized fractional Itô lemma and the related
'effective' Fokker-Planck equation. Later, in Section \ref{sec:Price model},
we arrive at the transition density function for a price process driven
by a gfBm with proportional drift (Black-Scholes type). In Section
\ref{sec:Applications}, we derive the Value-at-Risk and Expected
Shortfall, in addition to the close-form formula for European vanilla
contracts. At next, a \hyperref[sec:A-CEV-type-extension]{CEV-type exension}
is proposed computing European Call and Put prices. Finally, the main
conclussions are displayed.

\section{On the Generalized fractional Brownian motion}

Let $(a,b)\neq\left(0,0\right)\in\mathbb{R}^{2}$ and $H\in\left(0,1\right)$.
Be also $B_{t}^{H}$ a two-sided fBm, $\forall t\in\mathbb{R}_{+}$.
Then, the centered Gaussian process $Z_{t}^{H,a,b}$ is called gfBm
and defined by:

\begin{equation}
Z_{t}^{H,a,b}=aB_{t}^{H}+bB_{-t}^{H}\label{eq:Z}
\end{equation}

Depending on the values of $a,b$, and $H$ we could recover both
Bm, fBm, and sfBm. Its clear to see from the above definition that
$Z_{t}^{H,a,0}$ reduced the process to a fBm, $Z_{t}^{1/2,a,0}$
matches to a standard Bm, while $Z_{t}^{H,\frac{1}{\sqrt{2}},\frac{1}{\sqrt{2}}}=\left(B_{t}^{H}+B_{-t}^{H}\right)/\sqrt{2}=\xi_{t}^{H}$
corresponds to a definition of sfBm \citep[Eq.  2.1]{bojdecki2004sub}.

From Eq. (\ref{eq:Z}) we can also determine the covariance and variance
of the process.  Let $t,s\geq0$. Then the following results apply:

\[
E\left(Z_{t}^{H,a,b}\cdot Z_{s}^{H,a,b}\right)=\frac{\left(a+b\right)^{2}}{2}\left(s^{2H}+t^{2H}\right)-ab\left(s+t\right)^{2H}-\frac{a^{2}+b^{2}}{2}\left|t-s\right|^{2H}
\]

\[
E\left[\left(Z_{t}^{H,a,b}\right)^{2}\right]=\left[\left(a+b\right)^{2}-2^{2H}ab\right]t^{2H}
\]

From refs. \citep{zili2017generalized,zili2018generalized}, we observe
that the above covariance structure has a couple of implications in
the study of non-overlapping increments of the process (\ref{eq:Z}).
First, the increments are non-stationary unless $b=0$; i.e., in the
case of a fBm. Second, they are positively (resp. negative) correlated
if $H>1/2$ ($H<1/2$). Moreover, the magnitude of the autocorrelation
is stronger, weaker, or in between than the values provided by fBm
and sfBm, depending on the selected values of the triple $\left(H,a,b\right)$.
Indeed, we will have long-range dependence\footnote{Mathematically it can be determined analyzing the sum of the auto-covariance
for non-overlapping increments at different coupling times: $\sum_{n=1}^{\infty}E\left[\left(Z_{t+1}^{H,a,b}-Z_{t}^{H,a,b}\right)\cdot\left(Z_{t+n+1}^{H,a,b}-Z_{t+n}^{H,a,b}\right)\right]$,
for any integer $n$. If the sum diverges to infinity we are in the
presence of long-range dependence, while if the sum is finite we define
it as short-range dependence.} (slow decay on the autocorrelation) for any $a\neq b$ and $H>1/2$,
while for $H<1/2$ or $a=b$ and any $H$, short-range dependence
(high decay on the autocorrelation) appears. These features equipped
the gfBm with the ability to control the level of correlation between
the increments of the studied phenomena and the ductility to arise
stationary and non-stationary increments.

 On the other hand, since we have a bounded variance, we can derive
the related Ito-Wick formula following the approach of Nualart and
Taqqu \citep{Nualart2008}. Let $f=f\left(t,Z_{t}^{H,a,b}\right)\in C^{1,2}\left(\left(0,\infty\right)\times\mathbb{R}\right)$.
Then:

\begin{align}
\text{d}f & =\frac{\partial f}{\partial t}\text{d}t+\frac{\partial f}{\partial z}\text{d}Z_{t}^{H,a,b}+\frac{1}{2}\frac{\partial^{2}f}{\partial z^{2}}\text{d}\left\{ \left[\left(a+b\right)^{2}-2^{2H}ab\right]t^{2H}\right\} \nonumber \\
 & =\frac{\partial f}{\partial t}\text{d}t+\frac{\partial f}{\partial z}\text{d}Z_{t}^{H,a,b}+\frac{1}{2}\frac{\partial^{2}f}{\partial z^{2}}\left\{ \left[\left(a+b\right)^{2}-2^{2H}ab\right]\text{d}t^{2H}\right\} \nonumber \\
 & =\left\{ \frac{\partial f}{\text{\ensuremath{\partial t}}}+H\left[\left(a+b\right)^{2}-2^{2H}ab\right]t^{2H-1}\frac{\partial^{2}f}{\text{\ensuremath{\partial x^{2}}}}\right\} \text{d}t+\frac{\partial f}{\text{\ensuremath{\partial x}}}\text{d}Z_{t}^{H,a,b}\label{eq:Ito}
\end{align}

It's easy to show that for $a=b=1/\sqrt{2}$, the Eq. (\ref{eq:Ito})
corresponds to the sub-fractional It\^o's lemma \citep{yan2011ito}.
While, for $\left(a,b\right)=\left(1,0\right)$, we arrive at the
fractional It\^o 's lemma. \citep{bender2003ito}. The standard It\^o
calculus is recovered for $\left(a,b,H\right)=\left(1,0,1/2\right)$
or $\left(a,b,H\right)=\left(1/\sqrt{2},1/\sqrt{2},1/2\right)$.

In the same way, the previous result can be adapted to more general
processes driven by a gfBm. Let $y_{t}$ a stochastic process described
by the following stochastic differential equation (SDE):

\begin{equation}
\text{d}y_{t}=\mu\left(y_{t},t\right)\text{d}t+\sigma\left(y_{t},t\right)\text{d}Z_{t}^{H,a,b}\label{eq:dy}
\end{equation}

\noindent then, the transformation $g=g\left(y_{t},t\right)\in C^{2}\left(\mathbb{R}\right)$
follows:

\begin{equation}
\text{d}g\left(y_{t},t\right)=\left\{ \frac{\partial g}{\text{\ensuremath{\partial t}}}+\left(\frac{\partial f}{\text{\ensuremath{\partial y}}}\right)\mu+H\left[\left(a+b\right)^{2}-2^{2H}ab\right]\sigma^{2}t^{2H-1}\frac{\partial^{2}f}{\text{\ensuremath{\partial y^{2}}}}\right\} \text{d}t+\left(\frac{\partial f}{\text{\ensuremath{\partial y}}}\right)\sigma\text{d}Z_{t}^{H,a,b}\label{eq:dg}
\end{equation}

The last statement is also useful to derive the related ``effective''
Fokker-Planck equation that rules the transition density function
for the generic process (\ref{eq:dy}). Following \citep[Theorem 2.1]{araneda2021sub},
the transition density function $P$ is governed by:

\begin{equation}
\frac{\partial P}{\partial t}=H\left[\left(a+b\right)^{2}-2^{2H}ab\right]t^{2H-1}\frac{\partial^{2}}{\partial y^{2}}\left(\sigma^{2}P\right)-\frac{\partial}{\partial y}\left(\mu P\right)\label{eq:FPg}
\end{equation}

\section{Price modelling \label{sec:Price model}}

Let $S_{t}$ the price of a given stock. We will assume that it follows
the BS model driven by gmfBm. Then, under the real physical measure
$\mathbb{P}$, the evolution of the price is ruled by:

\begin{equation}
\text{d}S_{t}=\mu S_{t}\text{d}t+\sigma S_{t}\text{d}Z_{t}^{H,a,b}\label{eq:BS}
\end{equation}

\noindent being $\mu$ and $\sigma$ constant coefficients and $\text{d}Z_{t}^{H,a,b}$
represents the differential form of the process $Z_{t}^{H,a,b}$ defined
at Eq. (\ref{eq:Z}).

Applying the change of variable $x=\ln S-\mu t$ and the related It\^o's
calculus (Eq. \ref{eq:dg}), the new process $x$ obeys the following
SDE:

\begin{equation}
\text{d}x=-H\left[\left(a+b\right)^{2}-2^{2H}ab\right]t^{2H-1}\text{d}t+\sigma\text{d}Z_{t}^{H,a,b,c}\label{eq:x}
\end{equation}

From Eq. (\ref{eq:FPg}), the evolution of the transition density
function obeys the following PDE:

\begin{equation}
\frac{\partial P}{\partial t}=\sigma^{2}\left\{ H\left[\left(a+b\right)^{2}-2^{2H}ab\right]t^{2H-1}\right\} \left(\frac{\partial^{2}P}{\partial x^{2}}+\frac{\partial P}{\partial x}\right)\label{eq:FP}
\end{equation}

\noindent with initial condition $P(x,0)=\delta\left(x-x_{0}\right)$,
being $x_{0}$ the knowing value of $x$ at the inception time (which
implies that the density is concentrated at that point at the beginning).

By the time substitution

\[
\tau=\sigma^{2}\left[\left(a+b\right)^{2}-2^{2H}ab\right]t^{2H}
\]

\noindent  Eq. (\ref{eq:FP}) is transformed into a time-homogeneous
convection-diffusion equation:

\[
\frac{\partial P}{\partial\tau}=\frac{1}{2}\frac{\partial^{2}P}{\partial x^{2}}+\frac{1}{2}\frac{\partial P}{\partial x}
\]

After that, moving to a traveling frame of reference $w=x+\tau/2$,
we obtain:

\begin{equation}
\frac{\partial P}{\partial\tau}=\frac{1}{2}\frac{\partial^{2}P}{\partial w^{2}}\label{eq:diffBS}
\end{equation}

Eq. (\ref{eq:diffBS}) corresponds to an standard diffusion equation
with constant diffusion coefficient, with fundamental solution:

\begin{equation}
P(w,\tau)=\frac{1}{\sqrt{2\pi\tau}}\exp\left[-\frac{\left(w-w_{0}\right)^{2}}{2\tau}\right]\label{eq:P_w}
\end{equation}

\noindent where $P\left(w,0\right)=P\left(w_{0}\right)=\delta\left(w_{0}\right)$.
This initial condition is given by knowing the state of the asset
$S$ at the inception time; i.e., $S\left(t=0\right)=S_{0}=\text{e}^{x_{0}}=\text{e}^{w_{0}}$.

After replacing $w$ and $\tau$ as a function of $x$ and $t$, respectively;
we finally obtain:

\begin{multline}
P\left(\left.x,t\right|x_{0},0\right)=\left(2\pi\sigma^{2}\left[\left(a+b\right)^{2}-2^{2H}ab\right]t^{2H}\right)^{-\frac{1}{2}}\\
\times\exp\left[-\frac{\left(x-x_{0}+\frac{1}{2}\sigma^{2}\left[\left(a+b\right)^{2}-2^{2H}ab\right]t^{2H}\right)^{2}}{2\sigma^{2}\left[\left(a+b\right)^{2}-2^{2H}ab\right]t^{2H}}\right]\label{eq:Px}
\end{multline}

Moreover, the transition probability density function for the price
process (\ref{eq:BS}), at time $T$, is given by:

\begin{equation}
P\left(\left.S_{T},T\right|S_{0},0\right)=\frac{\text{e}^{-\frac{\left(\ln\left(\frac{S_{T}}{S_{0}}\right)+rT+\frac{1}{2}\sigma^{2}\left[\left(a+b\right)^{2}-2^{2H}ab\right]T^{2H}\right)^{2}}{2\sigma^{2}\left[\left(a+b\right)^{2}-2^{2H}ab\right]T^{2H};}}}{S_{T}\sqrt{2\pi\sigma^{2}\left[\left(a+b\right)^{2}-2^{2H}ab\right]T^{2H}}}\label{eq:P_S}
\end{equation}

From the above equation, we can show that the expected rate of return
is $\mu$:

\begin{eqnarray}
\mathbb{E^{P}}\left(S_{T}\right) & = & \int_{0}^{\infty}S_{T}P\left(S_{T},T\right)\text{d}S_{T}\nonumber \\
 & = & \int_{0}^{\infty}\text{e}^{x_{T}+\mu T}P\left(x_{T},T\right)\text{d}x_{T}\nonumber \\
 & = & \frac{\text{e}^{x_{0}+\mu T}}{\sqrt{2\pi\sigma^{2}\left[\left(a+b\right)^{2}-2^{2H}ab\right]T^{2H}}}\int_{0}^{\infty}\exp\left[-\frac{\left(x-x_{0}-\frac{1}{2}\sigma^{2}\left[\left(a+b\right)^{2}-2^{2H}ab\right]T^{2H}\right)^{2}}{\sigma^{2}\left[\left(a+b\right)^{2}-2^{2H}ab\right]T^{2H}}\right]\text{d}x_{T}\nonumber \\
 & = & \frac{S_{0}\text{e}^{\mu T}}{\sqrt{2\pi}}\int_{0}^{\infty}\exp\left[-\frac{u^{2}}{2}\right]\text{d}u\nonumber \\
 & = & S_{0}\text{e}^{\mu T}\label{eq:S0}
\end{eqnarray}

\section{Applications\label{sec:Applications}}

\subsection{Risk measures}

\subsubsection{Value-at-risk}

A very popular measure for tail-risk is the well-known \emph{Value-at-Risk},
VaR in short. It computes the minimum losses of a portfolio, for a
given investment horizon, considering a fixed-choose risk level $\alpha$.
This level is the quantile in the distribution of the future portfolio.
In other words, the $\alpha$-VaR set the probability equal to $\alpha$
for a loss greater than VaR. Typical values for $\alpha$ are 1\%
and 5\%. Thus, setting a portfolio from a long position in the asset
$S$ (purchased at the inception time), the $\alpha$-VaR at time
$T$ will be $S_{0}\text{e}^{rT}-V$, where $V$ is obtained from:

\begin{equation}
\int_{0}^{V}P\left(S_{T},T\right)\text{d}S_{T}=\alpha\label{eq:VAR}
\end{equation}

Then, assuming that the price is governed by the process \ref{eq:BS},
from Eqs. \ref{eq:Px}-\ref{eq:P_S}, we get:

\begin{eqnarray}
\alpha & = & \int_{0}^{V}P\left(S_{T},T\right)\text{d}S_{T}\nonumber \\
 & = & \int_{-\infty}^{\ln V-\mu t}P\left(x_{T},T\right)\text{d}x_{T}\nonumber \\
 & = & \frac{1}{\sqrt{2\pi\sigma^{2}\left[\left(a+b\right)^{2}-2^{2H}ab\right]T^{2H}}}\int_{-\infty}^{\ln V-\mu t}\exp\left[-\frac{\left(x-x_{0}-\frac{1}{2}\sigma^{2}\left[\left(a+b\right)^{2}-2^{2H}ab\right]T^{2H}\right)^{2}}{\sigma^{2}\left[\left(a+b\right)^{2}-2^{2H}ab\right]T^{2H}}\right]\nonumber \\
 & = & N\left(-\frac{\ln\left(\frac{S_{0}}{V}\right)+\mu T-\frac{1}{2}\sigma^{2}\left[\left(a+b\right)^{2}-2^{2H}ab\right]T^{2H}}{\sigma\sqrt{\left[\left(a+b\right)^{2}-2^{2H}ab\right]T^{2H}}}\right)\nonumber \\
 & = & N\left(-d_{2}^{a,b,H,V}\right)\label{eq:VarK}
\end{eqnarray}

\noindent where $N\left(\cdot\right)$ stands for the standard normal
cumulative density, and:

\[
d_{2}^{a,b,H,V}=\frac{\ln\left(\frac{S_{0}}{V}\right)+\mu T-\frac{1}{2}\sigma^{2}\left[\left(a+b\right)^{2}-2^{2H}ab\right]T^{2H}}{\sigma\sqrt{\left[\left(a+b\right)^{2}-2^{2H}ab\right]T^{2H}}}
\]
Then:

\begin{equation}
V=S_{0}\exp\left[\mu T-\frac{1}{2}\sigma^{2}\left[\left(a+b\right)^{2}-2^{2H}ab\right]T^{2H}+N^{-1}\left(\alpha\right)\sigma\sqrt{\left[\left(a+b\right)^{2}-2^{2H}ab\right]T^{2H}}\right]\label{eq:K}
\end{equation}

\noindent where $N^{-1}\left(\cdot\right)$ stands for the inverse
of the standard normal cumulative distribution function (quantile
function).

Finally, the Value-At-Risk for a given level $\alpha$ at time $T$
is:

\begin{multline*}
VaR_{T}^{\alpha}=S_{0}\text{e}^{rT}\\
-S_{0}\exp\left[\mu T-\frac{1}{2}\sigma^{2}\left[\left(a+b\right)^{2}-2^{2H}ab\right]T^{2H}+N^{-1}\left(\alpha\right)\sigma\sqrt{\left[\left(a+b\right)^{2}-2^{2H}ab\right]T^{2H}}\right]
\end{multline*}

\begin{figure}[t]
\subfloat[\label{fig:a-1}$\alpha=1\%$ and $H=0.25$]%
{\includegraphics[width=0.5\columnwidth]{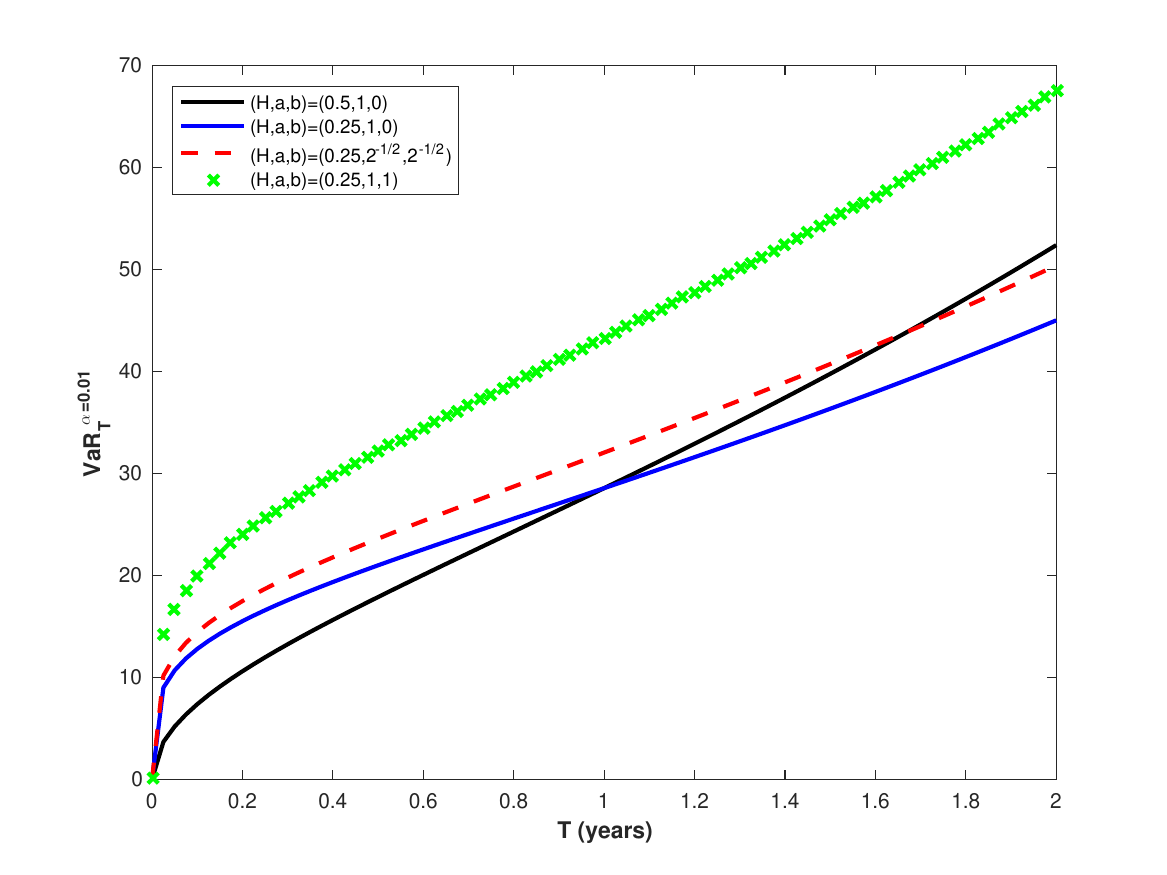}

}%
\subfloat[\label{fig:b-1}$\alpha=1\%$ and $H=0.75$]%
{\includegraphics[width=0.5\columnwidth]{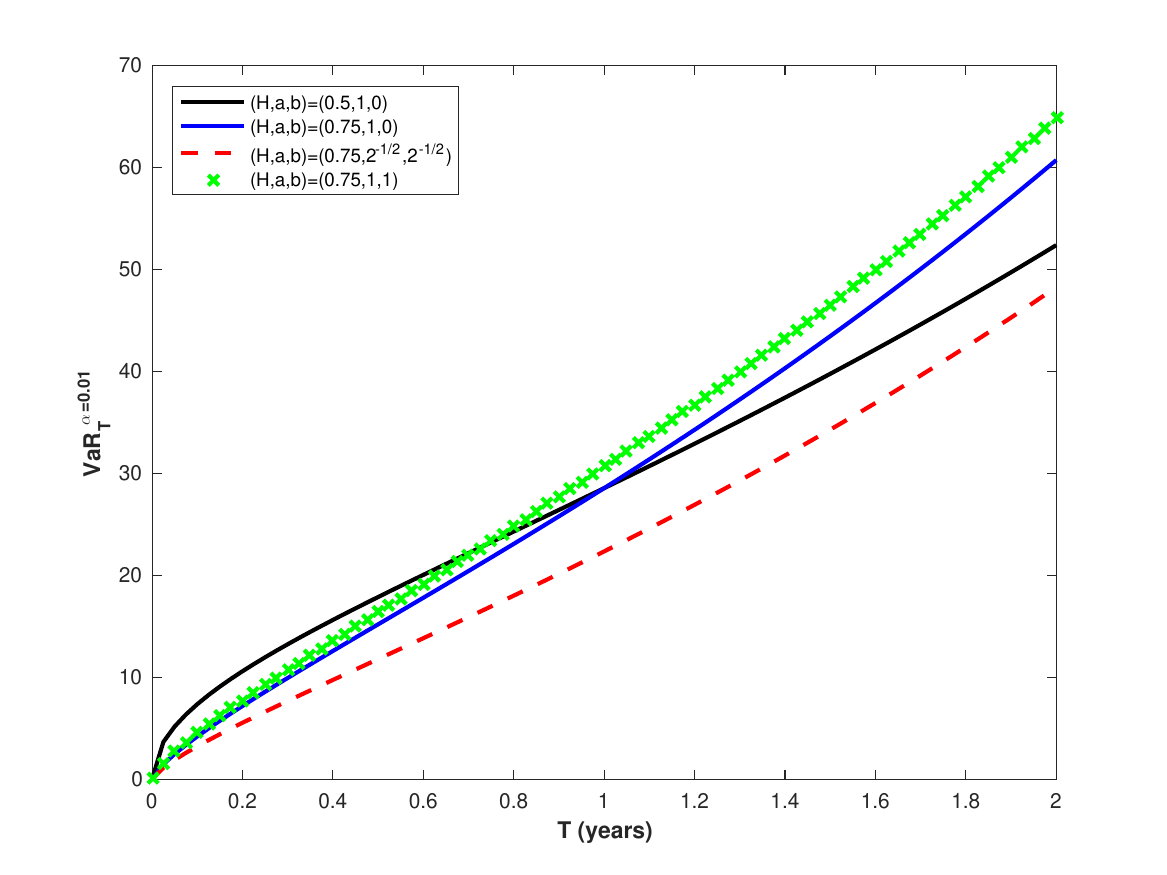}

}

\subfloat[\label{figc-1}$\alpha=5\%$ and $H=0.25$]%
{\includegraphics[width=0.5\columnwidth]{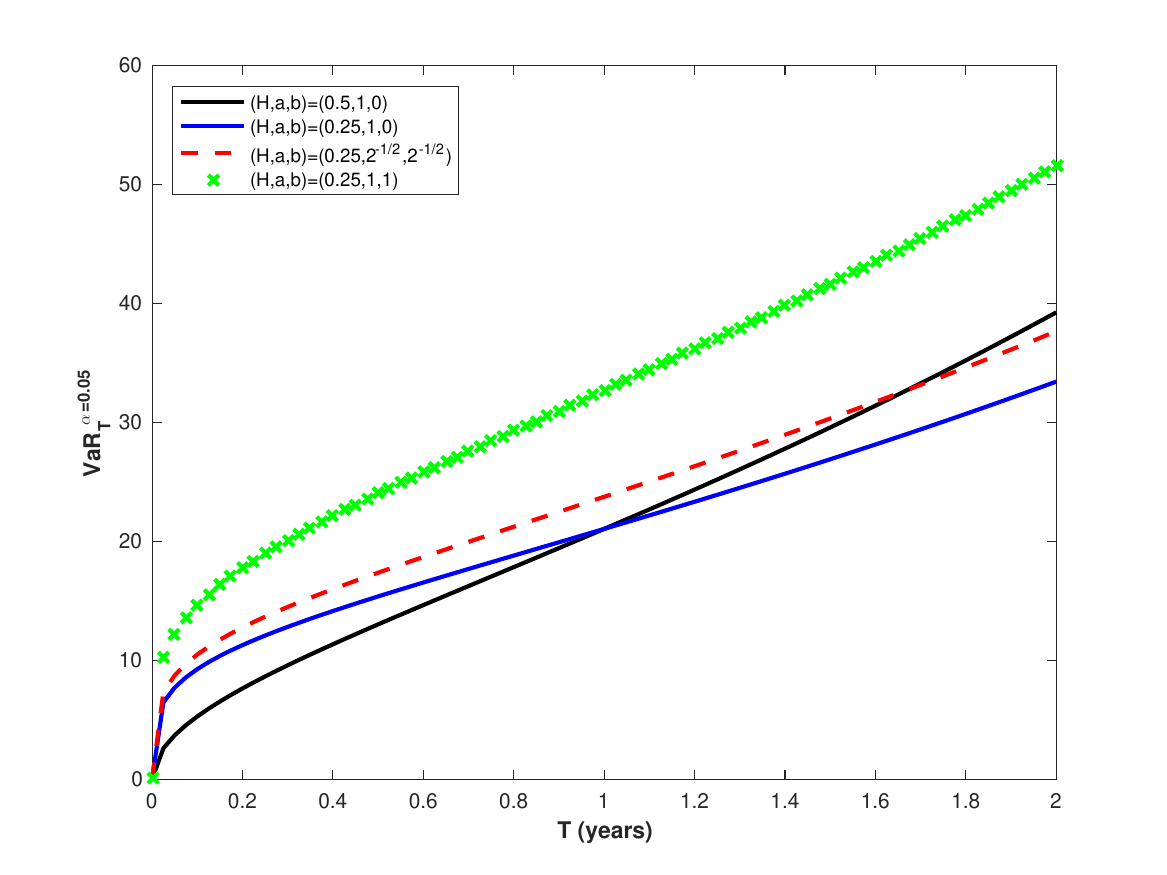}

}%
\subfloat[\label{figd-1}$\alpha=5\%$ and $H=0.75$]%
{\includegraphics[width=0.5\columnwidth]{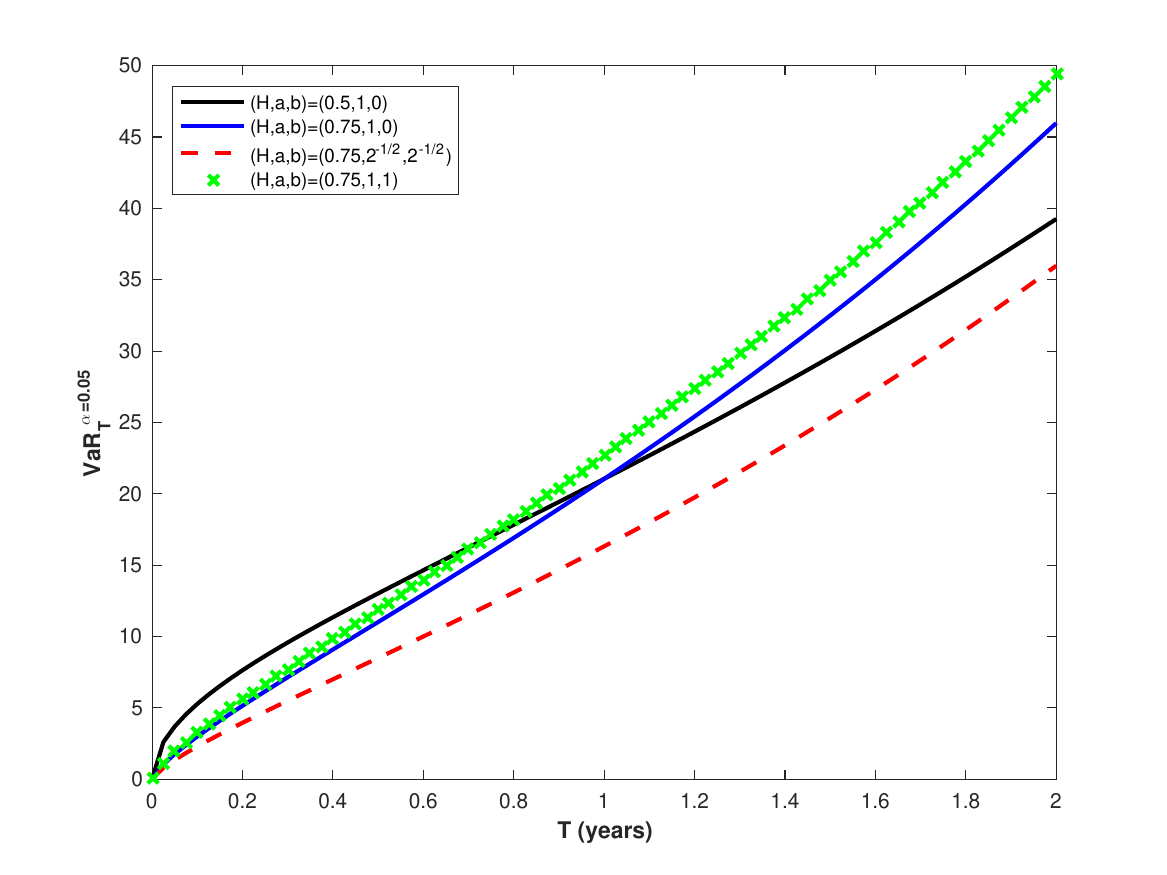}

}

\caption{\label{fig:Var}Value-at-Risk. $S_{0}=100$, $r=3\%$, and $\sigma=10\%$
were fixed as main parameters.}
\end{figure}

Figure \ref{fig:Var} shows the computations for Value-at-Risk as
a function of time, under the model \ref{eq:BS}. In Figs. \ref{fig:a-1}
and \ref{fig:b-1} we set $H=0.25$ and $H=0.75$, respectively, for
a quantile of 1\%. Otherwise, in Figs. \ref{figc-1}-\ref{figd-1},
the same values for $H$ are considered but $\alpha$ is increased
to 5\%. The standard Brownian motion, fractional and sub-fractional
schemes are ploted as particular cases of the proposed model. We can
infer from the plots that higher VaR are achieved for lower values
of $H$ and $b\neq0$. For the fractional case ($b=0$) this behavior
stands for $T<1$. In any case, a lower $H$ can increase the VaR
in short maturites. Moreover, according to the values of $a$ and
$b$, we can control the temporal structure of the VaR.

\subsubsection{Expected Shortfall}

A weakness of the VaR approach is it only reflects the minimum loss
on the portfolio for the quantile $\alpha$. An alternative approach
is the Expected Shortfall, a.k.a Conditional VaR \citep{Artzner1999}.
It's defined as the expected losses for a given VaR level . Mathematically,
the Expected Shortfall at level $\alpha$ and time $T$, $\text{ES}_{T}^{\alpha}$
is computed as:

\begin{eqnarray*}
\text{ES}_{T}^{\alpha} & = & S_{0}\text{e}^{rT}-\mathbb{\mathbb{E}}\left(\left.S_{T}\right|S_{T}<VaR_{T}^{\alpha}\right)\\
 & = & S_{0}\text{e}^{rT}-\frac{1}{\alpha}\int_{0}^{V}S_{T}P\left(S_{T},T\right)\text{d}S_{T}
\end{eqnarray*}

\noindent where $V$ is given by Eq. \ref{eq:K}, and:

\begin{eqnarray*}
\int_{0}^{V}S_{T}P\left(S_{T},T\right)\text{d}S_{T} & = & \int_{0}^{V}S_{T}P\left(S_{T},T\right)\text{d}S_{T}\\
 & = & \int_{-\infty}^{\ln V-\mu t}\text{e}^{x_{T}+\mu T}P\left(x_{T},T\right)\text{d}x_{T}\\
 & = & \frac{{\displaystyle \text{e}^{x_{0}+\mu T}\int_{-\infty}^{\ln V-\mu T}\text{e}^{-\frac{\left(x_{T}-x_{0}-\frac{1}{2}\sigma^{2}\left[\left(a+b\right)^{2}-2^{2H}ab\right]T^{2H}\right)^{2}}{2\sigma^{2}\left[\left(a+b\right)^{2}-2^{2H}ab\right]T^{2H}}}\text{d}x_{T}}}{\sqrt{2\pi\sigma^{2}\left[\left(a+b\right)^{2}-2^{2H}ab\right]T^{2H}}}\\
 & = & S_{0}\text{e}^{\mu T}N\left(-\frac{\ln\left(\frac{S_{0}}{V}\right)+\mu T+\frac{1}{2}\sigma^{2}\left[\left(a+b\right)^{2}-2^{2H}ab\right]T^{2H}}{\sigma\sqrt{\left[\left(a+b\right)^{2}-2^{2H}ab\right]T^{2H}}}\right)\\
 & = & S_{0}\text{e}^{\mu T}N\left(-d_{1}^{a,b,H,V}\right)
\end{eqnarray*}

\noindent with:

\[
d_{1}^{a,b,H}=\frac{\ln\left(\frac{S_{0}}{V}\right)+rT+\frac{1}{2}\sigma^{2}\left[\left(a+b\right)^{2}-2^{2H}ab\right]T^{2H}}{\sigma\sqrt{\left[\left(a+b\right)^{2}-2^{2H}ab\right]T^{2H}}}
\]

Moreover, since:

\[
d_{1}^{a,b,H,V}=d_{2}^{a,b,H,V}+\sigma\sqrt{\left[\left(a+b\right)^{2}-2^{2H}ab\right]T^{2H}}
\]

\noindent and by virtue of Eq. \ref{eq:VarK}, $-d_{1}^{a,b,H,V}=N^{-1}\left(\alpha\right)$,
we arrive at:

\[
\text{ES}_{T}^{\alpha}=S_{0}\text{e}^{rT}-S_{0}\frac{\text{e}^{\mu T}}{\alpha}N\left(N^{-1}\left(\alpha\right)-\sigma\sqrt{\left[\left(a+b\right)^{2}-2^{2H}ab\right]T^{2H}}\right)
\]

\begin{figure}[t]
\subfloat[\label{fig:a-1-1}$\alpha=1\%$ and $H=0.25$]%
{\includegraphics[width=0.5\columnwidth]{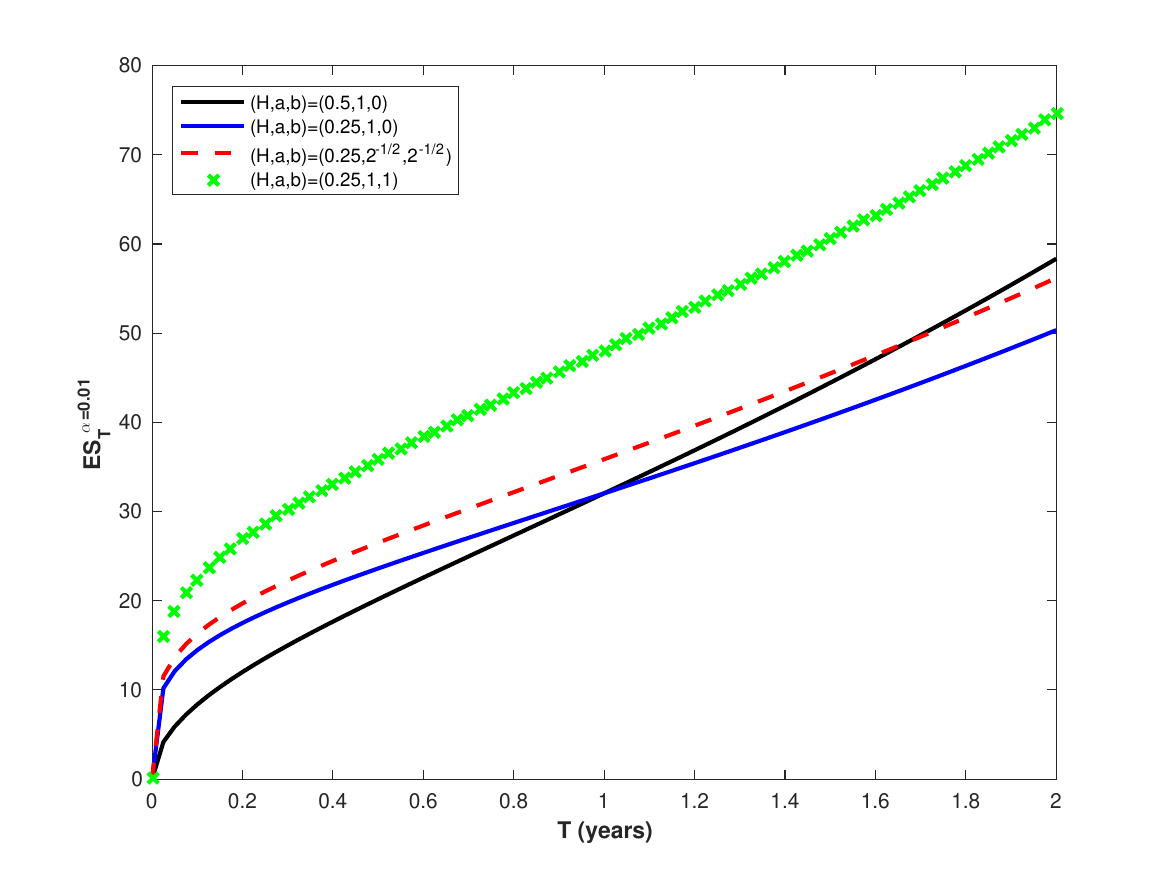}

}%
\subfloat[\label{fig:b-1-1}$\alpha=1\%$ and $H=0.75$]%
{\includegraphics[width=0.5\columnwidth]{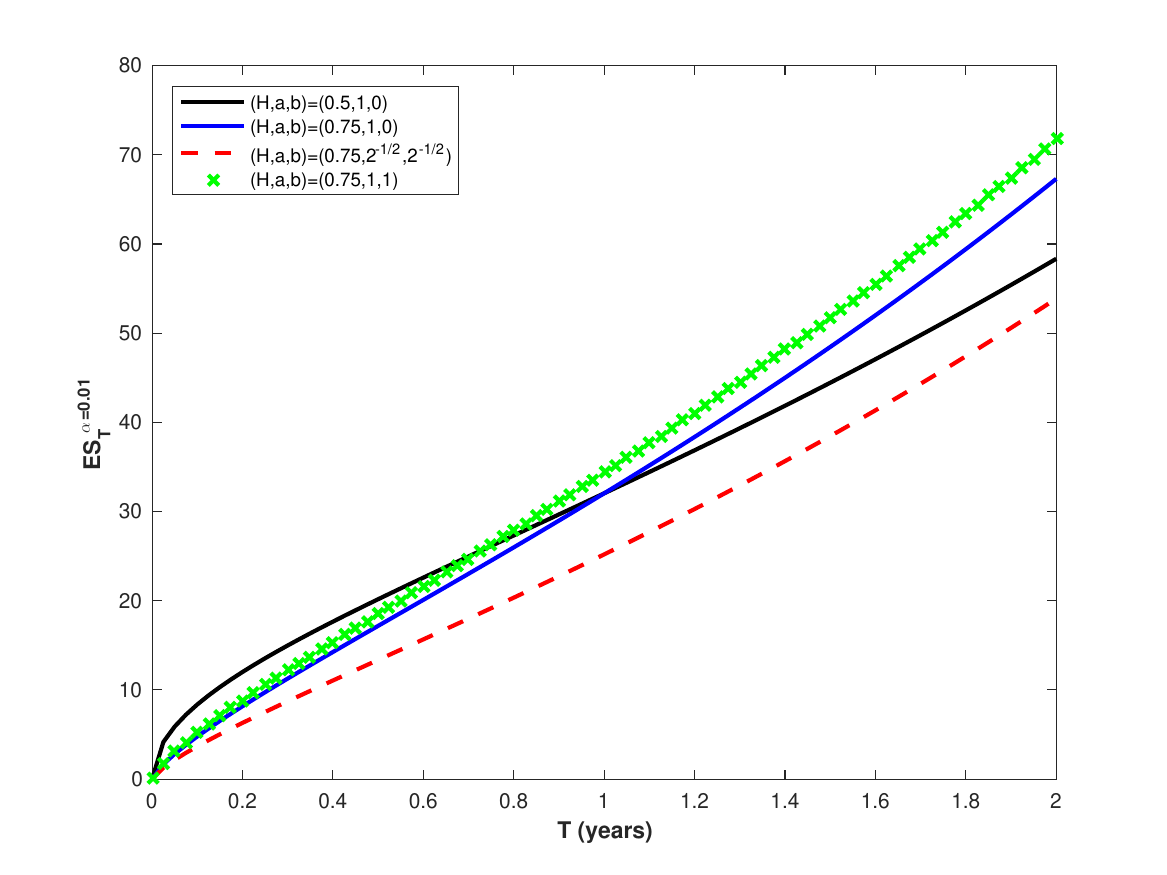}

}

\subfloat[\label{figc-1-1}$\alpha=5\%$ and $H=0.25$]%
{\includegraphics[width=0.5\columnwidth]{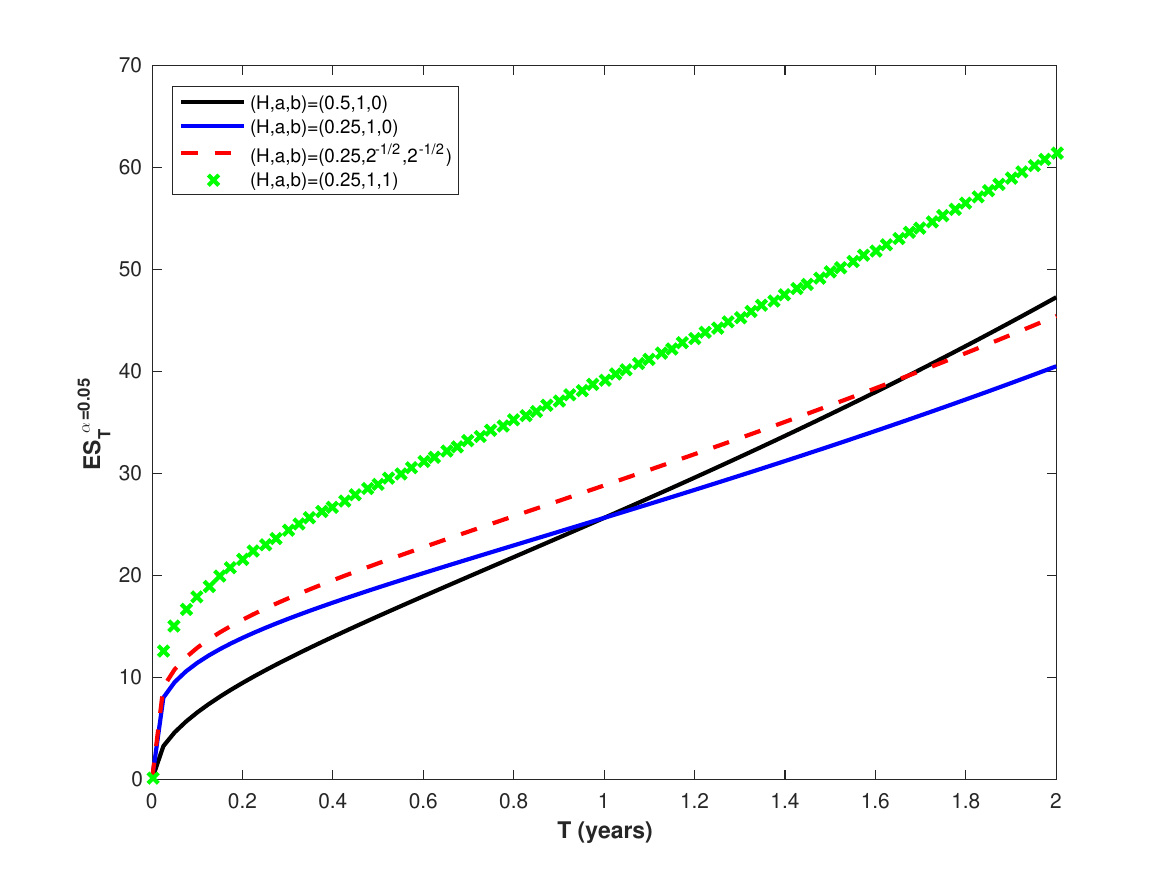}

}%
\subfloat[\label{figd-1-1}$\alpha=5\%$ and $H=0.75$]%
{\includegraphics[width=0.5\columnwidth]{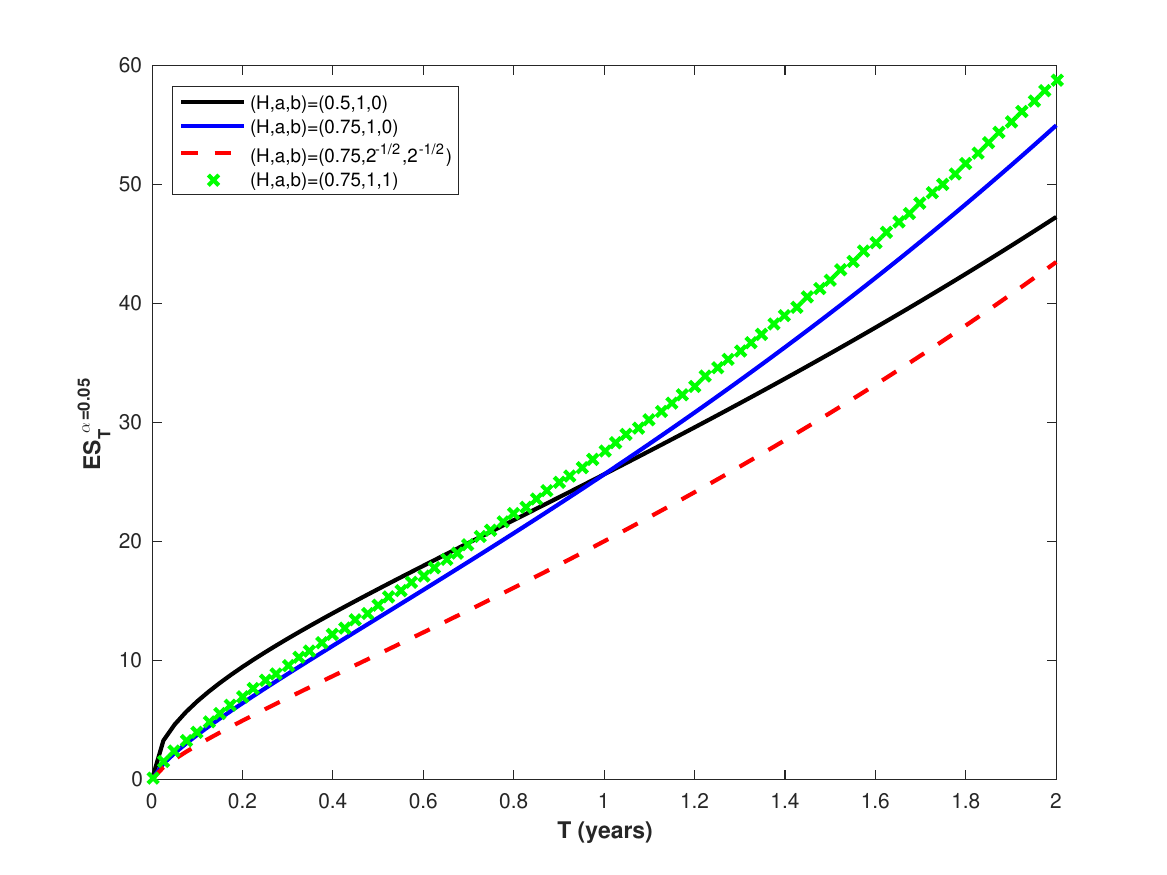}

}

\caption{\label{fig:ES}Expected Shortfall. $S_{0}=100$, $r=3\%$, and $\sigma=10\%$
were fixed as main parameters.}
\end{figure}

The Fig. \ref{fig:ES} reports the Expected Shortfall values related
to the VaR evaluates in Fig. \ref{fig:Var}.

\subsection{European option pricing\label{subsec:European-option-pricing}}

As pointed by Zili \citep{zili2017generalized}, gfBm is not a semimartingale
(unless $H=1/2$). This issue has two-fold: the existence of arbitrage
opportunities and the absence of the risk-neutral measure. Nonetheless,
we can skip any economical assumptions and obtain the option valuation
under the physical measure using the actuarial approach \citep{Bladt1998},
in a similar way how the fair value of an insurance contract is priced.
Since $\text{e}^{\mu T}=\frac{\mathbb{E^{P}}\left(S_{T}\right)}{S_{0}}$,
the fair premium for a vanilla European Call option with maturity
$T$ and exercise price $E$ is given by \citep{Bladt1998}:

\[
\text{CALL}\left(E,T\right)=\mathbb{E^{P}}\left[\left(\text{e}^{-\mu T}S_{T}-\text{e}^{-rT}E\right)^{+}\right]
\]

Given that:

\begin{eqnarray*}
\text{e}^{-\mu T}S_{T}>\text{e}^{-rT}E & \iff & \text{e}^{x_{T}}>\text{e}^{-rT}E\\
 & \iff & x_{T}>\ln E-rT
\end{eqnarray*}

\noindent we have:

\begin{eqnarray}
\text{CALL}\left(E,T\right) & = & \int_{\ln E-rT}^{\infty}\left(\text{e}^{x_{T}}-\text{e}^{-rT}E\right)P\left(x_{T},T\right)\text{d}x_{T}\nonumber \\
 & = & \int_{\ln E-rT}^{\infty}\text{e}^{x_{T}}P\left(x_{T},T\right)\text{d}x_{T}-E\text{e}^{-rT}\int_{\ln E-rT}^{\infty}P\left(x_{T},T\right)\text{d}x_{T}\label{eq:C}
\end{eqnarray}

Solving each one of the above integrals:

\begin{eqnarray*}
\int_{\ln E-rT}^{\infty}\text{e}^{x_{T}}P\left(x_{T},T\right)\text{d}x_{T} & = & \frac{{\displaystyle \int_{\ln E}^{\infty}\text{e}^{x_{T}}\text{e}^{-\frac{\left(x_{T}-x_{0}+\frac{1}{2}\sigma^{2}\left[\left(a+b\right)^{2}-2^{2H}ab\right]T^{2H}\right)^{2}}{2\sigma^{2}\left[\left(a+b\right)^{2}-2^{2H}ab\right]T^{2H}}}\text{d}x_{T}}}{\sqrt{2\pi\sigma^{2}\left[\left(a+b\right)^{2}-2^{2H}ab\right]T^{2H}}}\\
 & = & \frac{{\displaystyle \text{e}^{x_{0}}\int_{\ln E}^{\infty}\text{e}^{-\frac{\left(x_{T}-x_{0}-\frac{1}{2}\sigma^{2}\left[\left(a+b\right)^{2}-2^{2H}ab\right]T^{2H}\right)^{2}}{2\sigma^{2}\left[\left(a+b\right)^{2}-2^{2H}ab\right]T^{2H}}}\text{d}x_{T}}}{\sqrt{2\pi\sigma^{2}\left[\left(a+b\right)^{2}-2^{2H}ab\right]T^{2H}}}\\
 & = & S_{0}N\left(\frac{\ln\left(\frac{S_{0}}{E}\right)+rT+\frac{1}{2}\sigma^{2}\left[\left(a+b\right)^{2}-2^{2H}ab\right]T^{2H}}{\sigma\sqrt{\left[\left(a+b\right)^{2}-2^{2H}ab\right]T^{2H}}}\right)\\
 & = & S_{0}N\left(d_{1}^{a,b,H,E}\right)
\end{eqnarray*}

\begin{eqnarray*}
\int_{\ln E-rT}^{\infty}P\left(x_{T},T\right)\text{d}x_{T} & = & N\left(\frac{\ln\left(\frac{S_{0}}{E}\right)+rT-\frac{1}{2}\sigma^{2}\left[\left(a+b\right)^{2}-2^{2H}ab\right]T^{2H}}{\sigma\sqrt{\left[\left(a+b\right)^{2}-2^{2H}ab\right]T^{2H}}}\right)\\
 & = & N\left(d_{2}^{a,b,H,E}\right)
\end{eqnarray*}

Then,

\begin{equation}
\text{CALL}\left(E,T\right)=S_{0}N\left(d_{1}^{a,b,H,E}\right)-E\text{e}^{-rT}N\left(d_{2}^{a,b,H,E}\right)\label{eq:Price}
\end{equation}

It should be noted that the above formula has correspondence with
earlier results. If $\left(a,b\right)=\left(1,0\right)$, the pricing
yields for the fractional BS formula \citep{necula2002option}. While
for $a=b=1/\sqrt{2}$ and $c=1$, the valuation formula matches to
the sub-fractional BS formula \citep{tudor2008sub2}. Moreover, the
standard BS formula is recovered using $\left(a,b,H\right)=\left(1,0,1/2\right)$
or $\left(a,b,H\right)=\left(1/\sqrt{2},1/\sqrt{2},1/2\right)$.

Conversely, for European put options, the valuation goes to:

\begin{eqnarray}
\text{PUT}\left(E,T\right) & = & \mathbb{E^{P}}\left[\left.\left(\text{e}^{-rT}E-\text{e}^{-\mu T}S_{T}\right)^{+}\right|\text{e}^{-\mu T}S_{T}<\text{e}^{-rT}E\right]\nonumber \\
 & = & \int_{-\infty}^{\ln E-rT}\left(\text{e}^{-rT}E-\text{e}^{x_{T}}\right)P\left(x_{T},T\right)\text{d}x_{T}\nonumber \\
 & = & E\text{e}^{-rT}N\left(-d_{2}^{a,b,H}\right)-S_{0}N\left(-d_{1}^{a,b,H}\right)\label{eq:Put}
\end{eqnarray}

\begin{figure}[t]
\subfloat[\label{fig:a}$T=0.5$ and $H=0.25$]%
{\includegraphics[width=0.5\columnwidth]{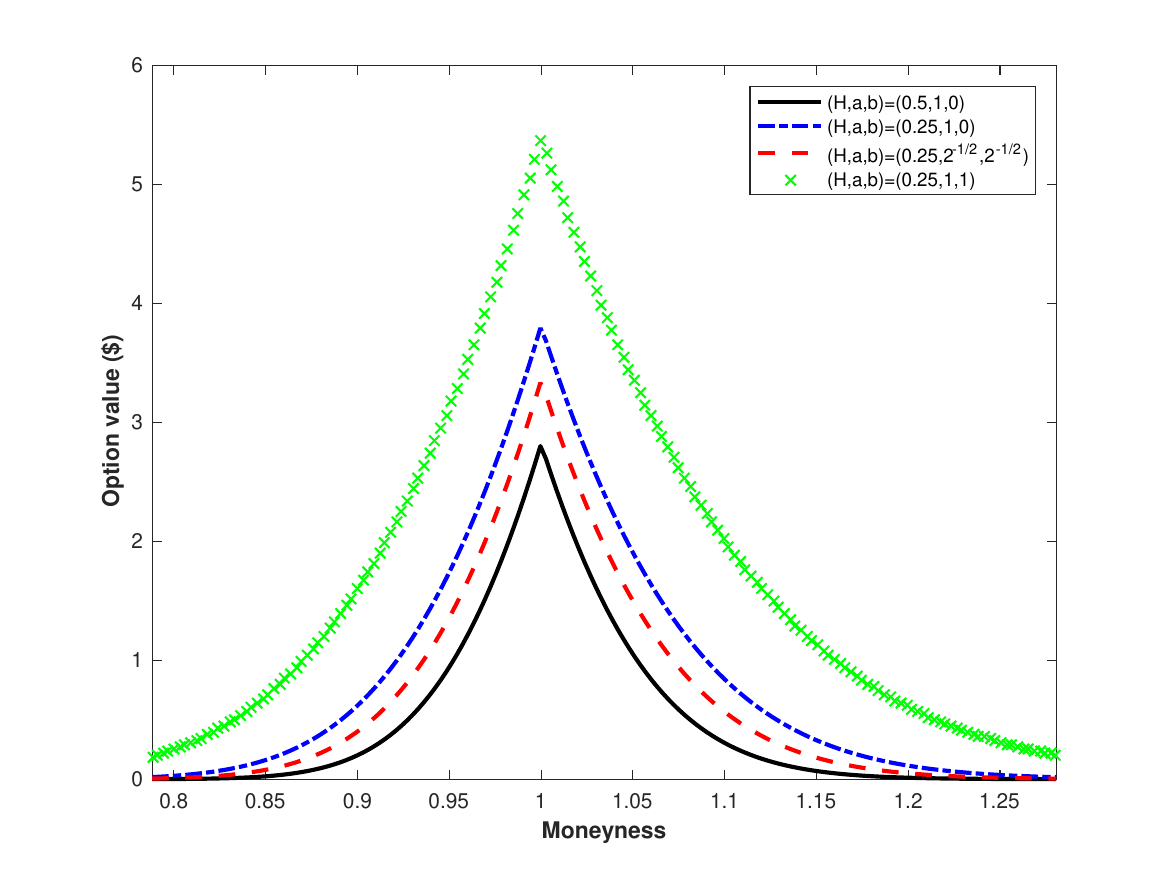}

}%
\subfloat[\label{fig:b}$T=0.5$ and $H=0.75$]%
{\includegraphics[width=0.5\columnwidth]{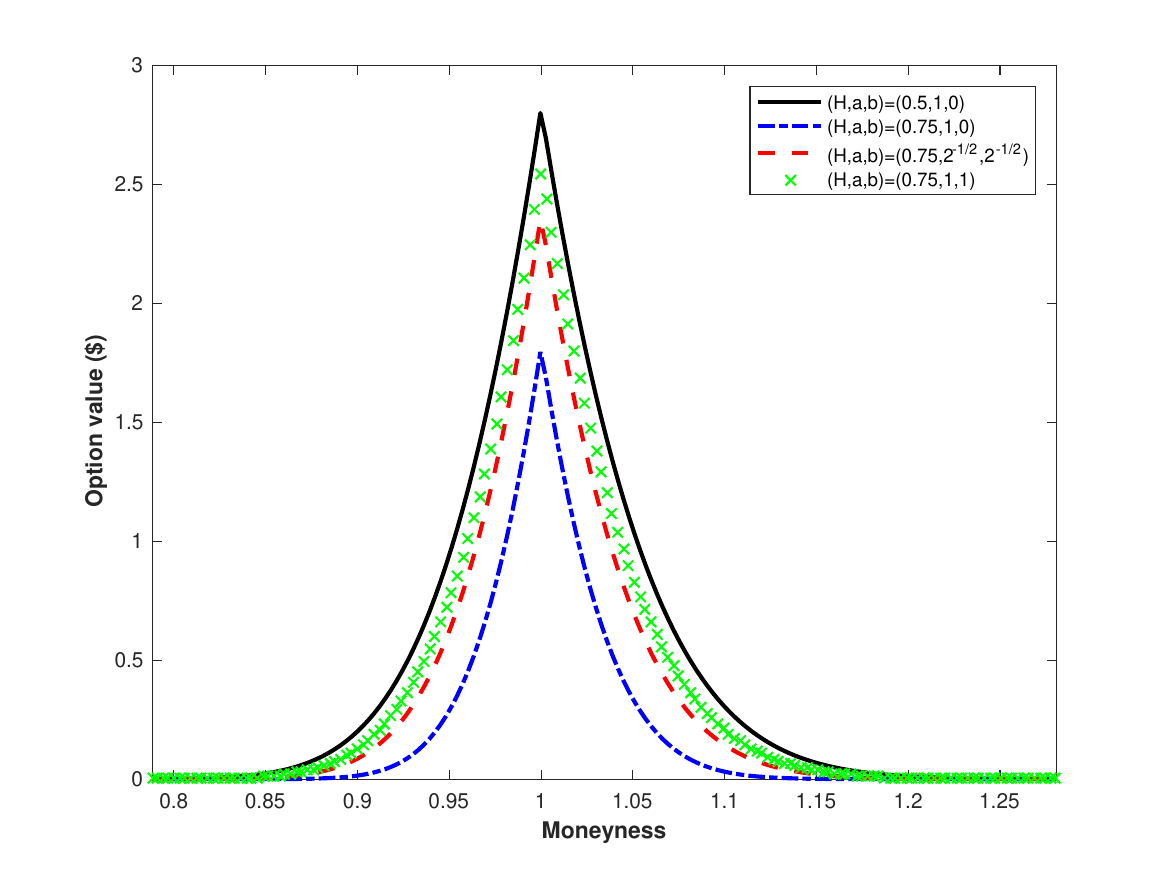}

}

\subfloat[\label{figc}$T=2$ and $H=0.25$]%
{\includegraphics[width=0.5\columnwidth]{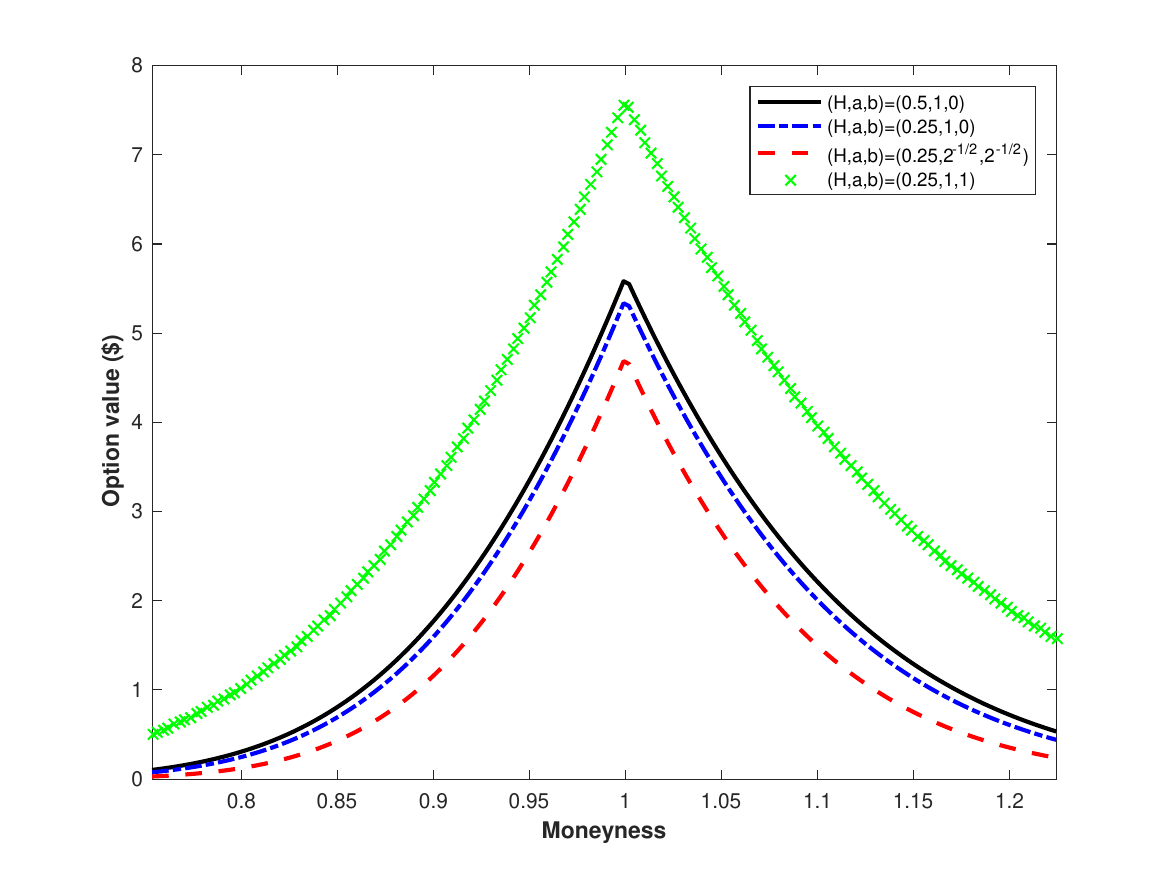}

}%
\subfloat[\label{figd}$T=2$ and $H=0.75$]%
{\includegraphics[width=0.5\columnwidth]{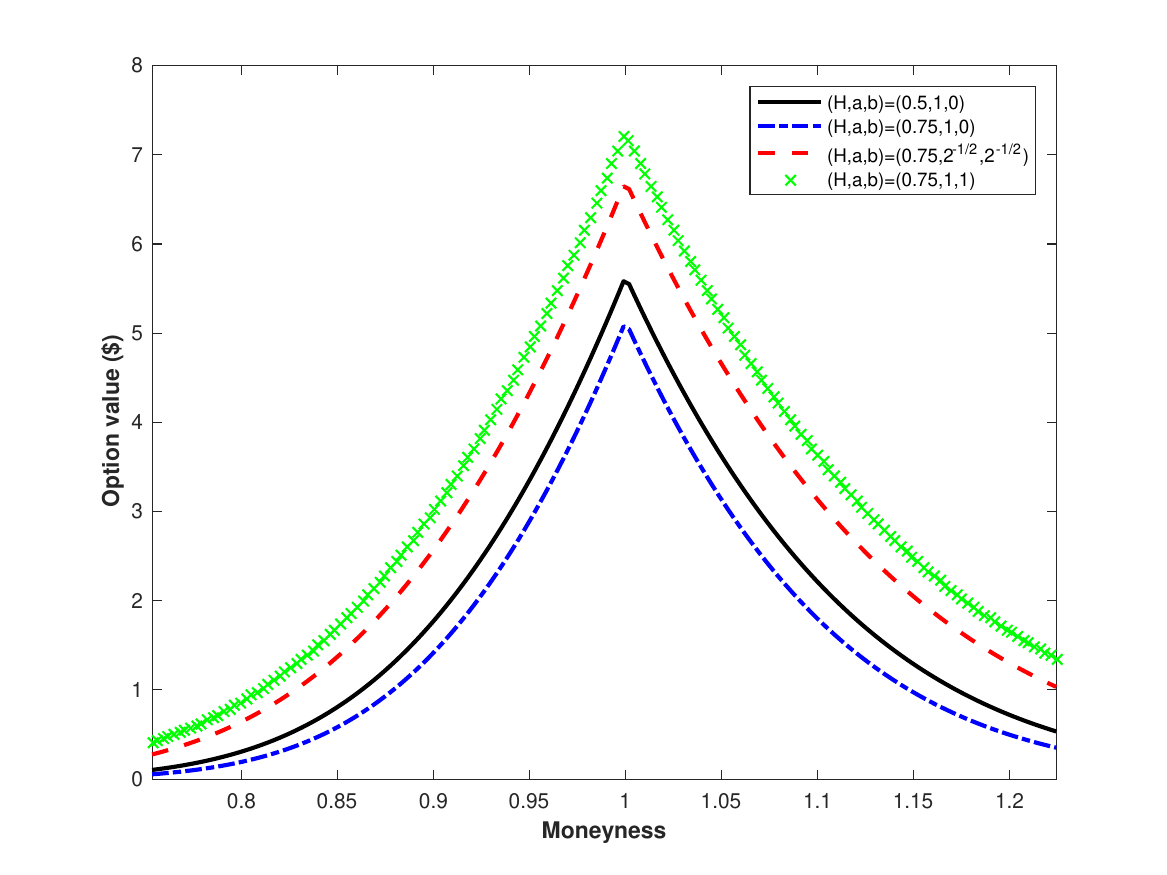}

}

\caption{\label{fig:Eupean-Option-prices}Out-of-the-money European Option
prices for the Black-Scholes model driven by gfBm. $S_{0}=100$, $r=3\%$,
and $\sigma=10\%$ were fixed as main parameters.}

\end{figure}

Fig. \ref{fig:Eupean-Option-prices} shows the European option prices,
from formulas \ref{eq:Price} and \ref{eq:Put}, for out-of-the-money
Calls and Puts as a function of moneyness ($Ee^{-rT}/S_{0}$); i.e.,
Put prices for $E<S_{0}e^{rt}$ and Call prices $E<S_{0}e^{rt}$.
The plot considers short (6 months in Figs. \ref{fig:a} and \ref{fig:b})
and long maturities (2 years in Figs. \ref{figc} and \ref{figd})
as well low ( Figs. \ref{fig:a} and \ref{figc}) and high (Figs.
\ref{fig:b} and \ref{figd}) Hurst values. The prices considers the
valuation under the proposed model with $a=b=1$ (green cross), but
also the reduced cases: the standard Black-Scholes formula (solid
black line), the fractional Black-Scholes model (blue dash-dot), and
the sub-fractional specification (red dashed line).

\section{A CEV-type extension\label{sec:A-CEV-type-extension}}

Lets consider now a further extension from Eq. (\ref{eq:BS}) where
the parameter $\sigma$ is allowed to be level-dependent; i.e., $\sigma=\sigma\left(S_{t}\right)$.
In particular, if we assume $\sigma=\sigma_{0}S_{t}^{\frac{\beta}{2}-1}$
with constant $\alpha$, we have a Constant Elasticity of Variance
structure\footnote{$\frac{\text{d}\left(v\right)/v}{\text{d}S/S}=\alpha-2$, $v=\text{Var\ensuremath{\left(\frac{\text{d}S_{t}}{S_{t}}\right)}}$},
as in the well-known CEV model proposed by Cox \citep{Cox1975,cox1996constant}.
Under generalized fractional CEV model, the price obeys the following
SDE:

\begin{equation}
\text{d}S_{t}=\mu S_{t}+\sigma_{0}S_{t}^{\frac{\beta}{2}}\text{d}M_{t}^{H,a,b}\label{eq:CEV-1}
\end{equation}

In the limit case, when the elasticity of variance goes to zero ($\beta=2$)
, the BS scheme is recovered. Moreover, if we restrict $\beta<2$
the leverage effect arises: inverse relationship between price and
volatility commonly observed in equity markets \citep{ballestra2016numerical}.
This direct relation price-volatility is capable to address the smile-skew
empirical fact \citep{eltit}. The rest of the papers considers $\beta<2$.

By the change of variables $y_{t}=S_{t}^{2-\beta}$, the process (\ref{eq:CEV-1})
goes to:

\begin{multline}
y_{t}=\left(2-\beta\right)\sigma\sqrt{y_{t}}\text{d}M_{t}^{H,a,b}+\left(2-\beta\right)\left\{ \mu y_{t}+\left(1-\beta\right)\sigma^{2}Ht^{2H-1}\left[\left(a+b\right)^{2}-2^{2H}ab\right]\right\} \mathrm{d}t\label{eq:GFCEV-1}
\end{multline}

According to, the evolution of the transition density function $P=P(y,t)$
obeys the following PDE:

\begin{equation}
\frac{\partial P}{\partial t}=\frac{\partial^{2}}{\partial y^{2}}\left[A(t)yP\right]-\frac{\partial}{\partial y}\left\{ \left[By+C(t)\right]P\right\} \label{eq:FP-1-1}
\end{equation}

\noindent with

\begin{eqnarray*}
A(t) & = & \left(2-\beta\right)^{2}\sigma^{2}Ht^{2H-1}\left[\left(a+b\right)^{2}-2^{2H}ab\right]\\
B & = & \left(2-\beta\right)\mu\\
C(t) & = & \frac{\left(1-\beta\right)}{\left(2-\beta\right)}A(t)
\end{eqnarray*}

\noindent and initial condition $P(y,0)=\delta\left(y-y_{0}\right)$
and $y_{0}=S_{0}^{2-\beta}$.

For the purpose of solve Eq. (\ref{eq:FP-1-1}), and taking advantage
of the constant ratio $\theta=\frac{A\left(t\right)}{C\left(t\right)}=\frac{\left(2-\alpha\right)}{\left(1-\alpha\right)}$,
we apply the following transformations:

\begin{gather}
\bar{y}=y_{t}\text{e}^{-t\cdot B(t)}\label{eq:bary}\\
\bar{P}=P\text{e}^{-t\cdot B(t)}\label{eq:barP}\\
\bar{t}={\displaystyle \frac{\phi(t)}{2}}\label{eq:bart}\\
\nonumber 
\end{gather}

\noindent with

\begin{eqnarray}
\phi\left(t\right) & = & \int_{0}^{t}A\left(\tilde{t}\right)\text{e}^{-B\tilde{t}}\text{d}\tilde{t}\nonumber \\
 &  & \left[\left(a+b\right)^{2}-2^{2H}ab\right]\frac{\sigma^{2}\left(2-\beta\right)^{2}}{2\left(2H+1\right)}\text{e}^{-\left(2-\alpha\right)\beta t}\\
 &  & \times t^{2H}\left\{ 2H+1+\text{e}^{\frac{1}{2}\left(2-\beta\right)\mu t}\left[\left(2-\beta\right)\mu t\right]^{-H}M_{H,H+1/2}\left[\left(2-\beta\right)\mu t\right]\right\} \label{eq:phi}
\end{eqnarray}

\noindent and $M_{\kappa,\upsilon}\left(l\right)$ the M-Whittaker
function.

Then, after replace in Eq. (\ref{eq:FP-1-1}), we obtain the following
parabolic equation with constant coefficients:

\begin{equation}
\frac{\partial\bar{P}}{\partial\bar{t}}=-\frac{\partial}{\partial\bar{y}}\left\{ 2\theta\bar{P}\right\} +\frac{\partial^{2}}{\partial\bar{y}^{2}}\left[2\bar{y}\bar{P}\right]\label{eq:FP_constant}
\end{equation}

The above expression matches the Fokker-Planck equation for a time-homogeneous
Feller process\footnote{In particular, a squared-Bessel process of order $2\theta$.}
\citep{feller1951two}, and in consequence, can be solved analytically.
Imposing an absorbing boundary condition at the zero-state (Dirichlet
boundary condition at the origin, $P_{x_{t}=0}^{abs}=0$) we arrive
at \citep{feller1951two,lindsay2012simulation}: 

\begin{equation}
\bar{P}\left(\left.\bar{y},\bar{t}\right|y_{0},0\right)=\frac{1}{2\bar{t}}\left(\frac{y_{0}}{\bar{y}}\right)^{\frac{1-\vartheta}{2}}\exp\left[-\frac{\left(\bar{y}+y_{0}\right)}{2\bar{t}}\right]I_{1/\left(2-\alpha\right)}\left(\frac{\sqrt{\bar{y}y_{0}}}{\bar{t}}\right)\label{eq:P_bar}
\end{equation}

\noindent where $I_{\nu}\left(\bullet\right)$ is the modified Bessel
function of the first kind of order $\nu$.

By the reason of Eqs. (\ref{eq:bary})-(\ref{eq:phi}), the transition
density for the process (\ref{eq:GFCEV-1}) is written as:

\begin{alignat}{1}
P\left(\left.S_{t},T\right|S_{0},0\right) & =P\left(\left.y_{T},T\right|y_{0},0\right)\frac{\partial y_{T}}{\partial S_{T}}\nonumber \\
 & =\left(2-\beta\right)k^{\frac{1}{2-\beta}}\left(lm^{1-2\beta}\right)^{\frac{1}{2\left(2-\beta\right)}}\text{e}^{-l-m}I_{1/\left(2-\beta\right)}\left(2\sqrt{lm}\right)\label{eq:P1}
\end{alignat}

\noindent with

\begin{eqnarray}
k & = & \left[\phi\left(T\right)/2\right]^{-1},\label{eq:k}\\
l & = & kS_{0}^{2-\beta}\text{e}^{u\left(2-\beta\right)T}\label{eq:l}\\
m & = & kS_{T}^{2-\beta}\label{eq:m}
\end{eqnarray}

Further, with the help of the density (\ref{eq:P1}), and taking conditional
expectations for the price process with $\alpha<2$, we can arrive
at:

\begin{eqnarray*}
\mathbb{E^{P}}\left(S_{T}\right) & = & \int_{0}^{\infty}S_{T}P\left(S_{T},T\right)\text{d}S_{T}\\
 & = & \int_{0}^{\infty}\left(\frac{m}{k}\right)^{\frac{1}{2-\beta}}\left(\frac{l}{m}\right)^{\frac{1}{2\left(2-\beta\right)}}\text{e}^{-l-m}I_{1/\left(2-\beta\right)}\left(2\sqrt{lm}\right)\text{d}m\\
 & = & \int_{0}^{\infty}\left(\frac{l}{k}\right)^{\frac{1}{2-\alpha}}\left(\frac{m}{l}\right)^{\frac{1}{2\left(2-\beta\right)}}\text{e}^{-l-m}I_{1/\left(2-\beta\right)}\left(2\sqrt{lm}\right)\text{d}m\\
 & = & S_{0}\text{e}^{uT}\int_{0}^{\infty}\underbrace{\left(\frac{2m}{2l}\right)^{\frac{1}{2\left(2-\beta\right)}}\text{e}^{-\frac{2l+2m}{2}}I_{1/\left(2-\beta\right)}\left(\sqrt{\left(2l\right)\left(2m\right)}\right)}_{{\displaystyle f\left(2m;2+\frac{2}{2-\beta},2l\right)}}\text{d}m\\
 & = & S_{0}\text{e}^{uT}
\end{eqnarray*}

The last step is achieved taking in account that $f\left(u;\nu,\lambda\right)=\left(u/\lambda\right)^{\frac{\nu-2}{4}}\text{e}^{-\left(x+\lambda\right)/2}I_{\nu}\left(\sqrt{x\lambda}\right)$
matches to the density function for the non-central chi-square distribution
with non-centrality parameter $\lambda$ and $\nu$degree of freedoms,
$u\sim\chi_{\nu}^{2}\left(\lambda\right)$, and its complementary
function defined as:

\[
\int_{a}^{\infty}f\left(u;\nu,\lambda\right)\text{d}u=Q\left(a;\nu,\lambda\right)
\]

\noindent where $Q\left(0;\nu,\lambda\right)=1$.

Applying the arguments discussed in section \ref{sec:Price model},
for $\alpha<2$, and since $S_{T}\text{e}^{-\mu T}>E\text{e}^{-rT}\iff S_{T}>E\text{e}^{\left(\mu-r\right)t}$,
the pricing for an European Call option is computed as:

\begin{eqnarray}
\text{CALL}\left(E,T\right) & = & \mathbb{E^{P}}\left[\left(\text{e}^{-\mu T}S_{T}-\text{e}^{-rT}E\right)^{+}\right]\nonumber \\
 & = & \int_{E\text{e}^{\mu T-rT}}^{\infty}\left(\text{e}^{-\mu T}S_{T}-\text{e}^{-rT}E\right)P\left(S_{T},T\right)\text{d}S_{T}\label{eq:C_cev0}
\end{eqnarray}

\noindent where, 

\begin{eqnarray*}
\int_{E\text{e}^{\mu T-rT}}^{\infty}\text{e}^{-\mu T}S_{T}P\left(S_{T},T\right)\text{d}S_{T} & = & \text{e}^{-\mu T}\int_{kE^{2-\beta}\text{e}^{\left(2-\beta\right)\left(u-r\right)T}}^{\infty}\left(\frac{m}{k}\right)^{\frac{1}{2-\beta}}\left(\frac{l}{m}\right)^{\frac{1}{2\left(2-\beta\right)}}\\
 &  & \times\text{e}^{-l-m}I_{1/\left(2-\beta\right)}\left(2\sqrt{lm}\right)\text{d}m\\
 & = & S_{0}\int_{kE^{2-\beta}\text{e}^{\left(2-\beta\right)\left(u-r\right)T}}^{\infty}\left(\frac{2m}{2l}\right)^{\frac{1}{2\left(2-\beta\right)}}\text{e}^{-\frac{2l+2m}{2}}\\
 &  & \times I_{1/\left(2-\beta\right)}\left(\sqrt{\left(2l\right)\left(2m\right)}\right)\text{d}m\\
 & = & S_{0}\int_{F}^{\infty}{\displaystyle f\left(2m;2+\frac{2}{2-\beta},2l\right)}\text{d}m\\
 & = & S_{0}Q\left(2F;2+\frac{2}{2-\beta},2l\right)
\end{eqnarray*}

\noindent being $F=kE^{2-\beta}\text{e}^{\left(2-\beta\right)\left(u-r\right)T}$.
Moreover,

\begin{eqnarray*}
\int_{E\text{e}^{\mu T-rT}}^{\infty}E\text{e}^{-rT}P\left(S_{T},T\right)\text{d}S_{T} & = & E\text{e}^{-rT}\int_{F}^{\infty}{\displaystyle f\left(2l;2+\frac{2}{2-\beta},2m\right)}\text{d}m
\end{eqnarray*}

Making use of the property \citep{schroder1989computing}:
\[
\int_{a}^{\infty}{\displaystyle f\left(2u;2v,2\lambda\right)\text{d}\lambda}=1-Q\left(2u;2v-2,2a\right)
\]

\noindent we get:

\[
\int_{F}^{\infty}{\displaystyle f\left(2l;2+\frac{2}{2-\beta},2m\right)}\text{d}m=1-Q\left(2l;\frac{2}{2-\beta},2F\right)
\]

Then, after replace on Eq. \ref{eq:C_cev0} the Call valuation goes
to:

\begin{equation}
\text{CALL}\left(E,T\right)=S_{0}Q\left(2F;2+\frac{2}{2-\beta},2l\right)-E\text{e}^{-rT}\left[1-Q\left(2l;\frac{2}{2-\beta},2F\right)\right]\label{eq:CALL-CEV-1}
\end{equation}

Analogously, for an European put, we arrive at:

\[
\text{PUT}\left(E,T\right)=E\text{e}^{-rT}Q\left(2l;\frac{2}{2-\alpha},2F\right)-S_{0}\left[1-Q\left(2F;2+\frac{2}{2-\alpha},2l\right)\right]
\]

\section{Summary\label{sec:conclussions}}

In this paper we have used the gfBm as the driven process in the price
fluctuations modeling. By means of the related both Ito calculus and
effective Fokker-Planck equation, we have obtained the analytical
valuation for risk measures as Value-at-Risk and Expected Shortfall,
as well the close-form European Call pricing formulas for the generalized
fractional BS model. In addition, the CEV extension is discussed derived
its transition density and option pricing. Since that gmfBm is a generalization
for both Bm, fBm, and sfBm; we can recovery the the classical results
for the BS model, but also their fractional and sub-fractional extensions.
The study of exotic options as well the inclusion of stochastic volatility
models is pending to address in the future. 

\section*{}

\bibliographystyle{13_Users_axelaraneda_Desktop_Research_fBM_ws-fnl}
\addcontentsline{toc}{section}{\refname}\bibliography{12_Users_axelaraneda_Desktop_Research_fBM_Subfractional2}

\end{document}